\documentclass[%
 reprint,
 amsmath,amssymb,
 aps,
]{revtex4-2}

\usepackage{graphicx}% Include figure files
\usepackage{dcolumn}% Align table columns on decimal point
\usepackage{bm}% bold math
\begin{document}

\preprint{APS/123-QED}

\title{Tunable nonlinear coherent perfect absorption and reflection in cavity QED}% Force line breaks with \\

\author{Miaodi Guo}%
 \email{guomiaodi@xatu.edu.cn}
\affiliation{School of Sciences, Xi'an Technological University, Xi'an 710021, China}

\date{\today}% It is always \today, today,

\begin{abstract}
We propose and analyze a scheme for realizing tunable coherent perfect absorption (CPA) and reflection (CPR) in a three-level $\Lambda$-type atom-cavity system. With EIT-type interference induced by a coherent coupling laser, the scheme provides a new method of attaining tunable near CPR at two-photon resonance. By different coupling strength, the system can be excited from linear CPA regime into nonlinear even bistable CPA regime without changing the  frequency of the probe field, where the hysteresis cycle of the bistability is controllable. In addition, the CPA regime can be transferred into the near CPR regime with controllable atomic coherence and linear absorption induced by an incoherent pump field. Our work suggests a mechanism for manipulation on extreme absorption ($0\%$ and $100\%$), which can potentially be applied in logical gates and optical switches for coherent optical computing and communication. 
\end{abstract}

\maketitle

%\tableofcontents

\section{Introduction}
The ability to modify the linear absorption, dispersion, and nonlinearity of atomic media is crucial in various areas of research, e.g., all-optical switching, optical probe, optical memories, and logical gates \cite{Borges2016,Hao2019,Gao2019,Ann2020}. Coherent perfect absorption (CPA), as the realization of extreme absorption, arises from the time-reversed process of lasing at threshold which corresponds to a zero eigenvalue of the scattering matrix \cite{Chong2010}. It has been realized in solid-state Fabry-Perot structure \cite{Wan2011}, metamaterials \cite{Kang2018}, semiconductors \cite{Horng2020}, and whispering gallery mode (WGM) microcavities \cite{Wang2021}. When the system is driven by a quantum field, e.g. squeezed light, the quantum CPA can be realized \cite{Hardal2019}. These studies of CPA are under the linear absorption regime, in which the linear dielectrics are used as the absorber. Recently studies show that the nonlinear CPA can also be realized when the nonlinear dielectrics are considered in the solid-state devices such as epsilon-near-zero plasmonic waveguides and nonlinear metasurfaces \cite{Li:18,Alaee:20}. 

However, the absorption regime in solid-state materials is generally invariable when the material is determined. In cavity quantum electrodynamics (CQED) system, when the first-order polariton states are considered and the multiphoton excitation of the higher-order polariton states are excluded, the CPA is attained under linear excitation regime \cite{Agarwal2015}. In the strong coupling regime of CQED, or a weak coupling regime with a second-order nonlinear crystal (SOC) in the atom-cavity system, the linear and nonlinear excitation regimes are both included, as a result, the CPA can be tuned from the linear regime into the nonlinear regime \cite{Wang2017a,Xiong2020}. In addition, varying the frequency of the input probe field, the normally nonlinear CPA regime can also be excited into a bistable CPA regime, where the optical bistability (OB) that also plays an important role in the all-optical processing elements \cite{PRL108/263905,Sheng13,Nozaki2012,DelBino:21} is attained in a two-level atom-cavity system \cite{Xiong2020,Agarwal2016}. Even though, the excitation regime is determined for a certain frequency of the probe field in a two-level system \cite{Wei2018}. In comparison with that, the CPA in a multi-level system can be exchanged among the linear, normally nonlinear, and bistable regime at a certain frequency because of the modifying absorption, dispersion, and nonlinearity by atom coherence and quantum interference \cite{Jafarzadeh2014,Wu2016}. 

In this study, we analyze the tunable CPA and CPR in a three-level atom-cavity system. Comparing with the two-level system, this scheme provides a new method of modifying the frequency range of CPA and the excitation regime of the system, and a method of attaining four-frequency CPA and the near CPR state. In previous study of CPA, the influence of incoherent process on CPA condition that changes the linear absorption of the probe fields, has not been analyzed. In our scheme, an incoherent pump field is applied to modify the population at atomic levels, which will induce the controllable linear absorption by different pumping rates. That will further induce the transfer between the bistable CPA and the linear CPA, and also form the exchange of CPA and near CPR. Generally, the normally nonlinear CPA and the bistable CPA can be attained at the different frequencies of an input probe field. However, with a coherent control field coupling one of the ground states and the excited state of the atoms, two types of CPA can be attained at the same frequency. In addition, the near CPR and the four-frequency CPA can be tunable by the coherent coupling field or the incoherent pump field.

\section{Theoretical analysis}
\begin{figure}[!hbt]
	\centering
	\includegraphics[width=4cm]{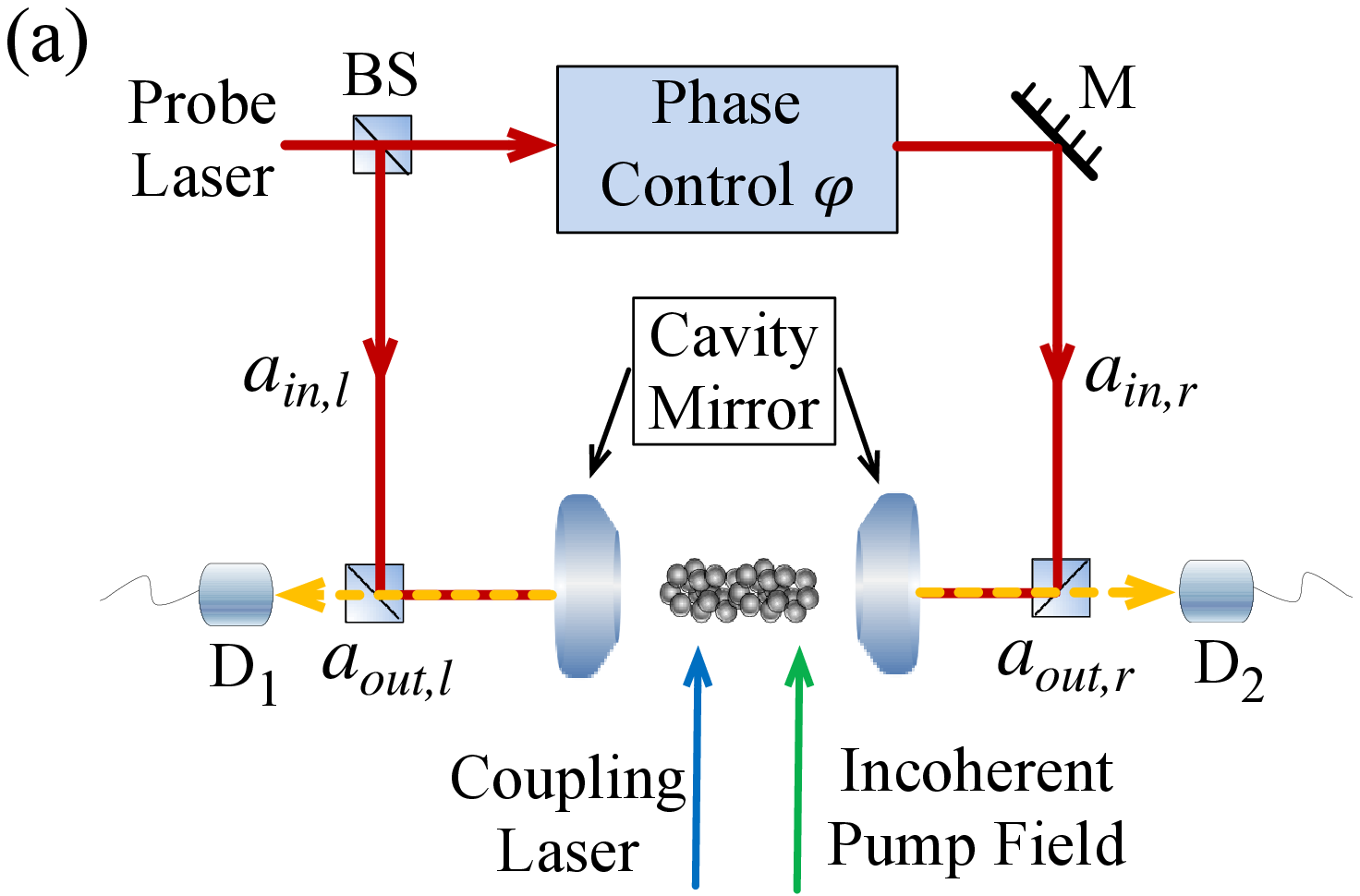}
	\includegraphics[width=4cm]{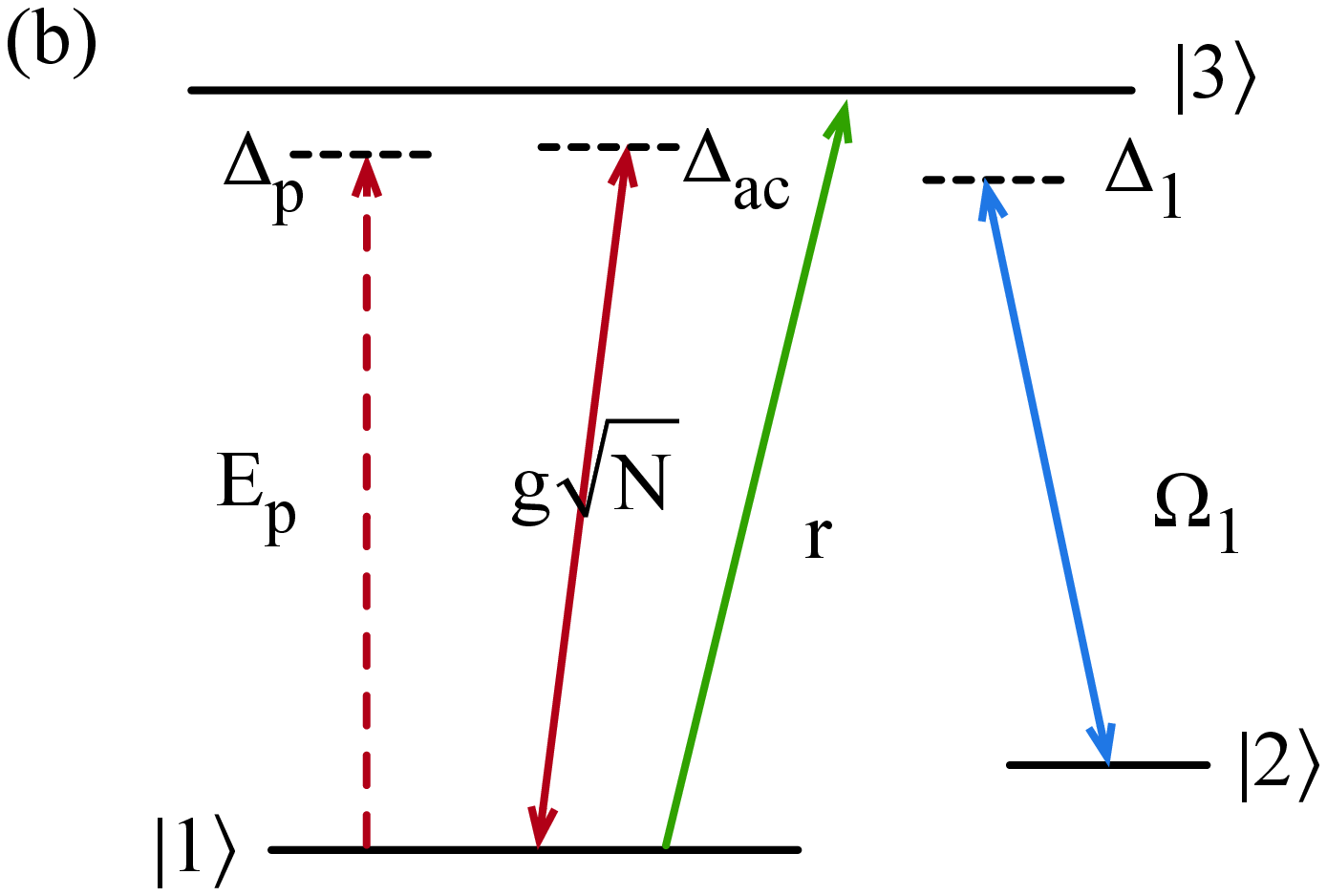}
	\caption{(a) Schematic diagram of a two-sided cavity filling with the (b) three-level atoms.}\label{fig-system}
\end{figure}

%\subsection{Proposed system}
The scheme proposed here is depicted in Fig. \ref{fig-system}(a). For simplicity, some cold atoms (the energy levels are shown in Fig. \ref{fig-system}(b)) are trapped in a single-mode cavity. A probe laser is split into two beams ($a_{in,l}$ and $a_{in,r}$) through a beam splitter (BS) that are injected into the cavity along the cavity axis with a relative phase ($\varphi$) and drive the atomic transition $|1\rangle\rightarrow|3\rangle$. An incoherent pump field is applied for population pumped from atomic energy level $|1\rangle$ to $|3\rangle$. A coupling laser is injected into the cavity perpendicularly to the cavity axis, which couples atomic energy levels $|2\rangle$ and $|3\rangle$. And two detectors are employed to measure the output fields ($a_{out,l}$ and $a_{out,r}$). For implementation, an example of such an atomic system is $^{87}$Rb for which $|1\rangle\equiv(5S_{1/2}, F=1)$, $|2\rangle\equiv(5S_{1/2}, F=2)$, and $|3\rangle\equiv(5P_{3/2}, F=2)$. And a laser of 780 nm can be used as the probe and coupling lasers. 

Under the rotating wave approximation, the off-resonant coupling terms are omitted, and the Hamiltonian of the system is given as follows (the incoherent pump mechanism can be ignored here) \cite{Sawant2016a},
\begin{equation}
\begin{split}
H=&-\hbar\{(\Delta_{p}-\Delta_{ac})a^{\dag}a+\sum_{j=1}^{N}[(\Delta_{p}-\Delta_{1})\sigma_{22}^{j}\\
&+\Delta_{p}\sigma_{33}^{j}+ga^{\dag}\sigma_{13}^{j}+\Omega_{1}\sigma_{23}^{j}]\}+H.C.,\label{equ-Hamiltonian}
\end{split}
\end{equation}
where $\hbar$ is the reduced Planck constant, $N$ is the number of the atoms inside the cavity, $H.C.$ denotes the Hermitian conjugate, $\Delta_{p}=\omega_{p}-\omega_{31}$ ($\Delta_{ac}=\omega_{c}-\omega_{31}$) is the frequency detuning of the probe laser (the cavity mode) and the atomic transition $|1\rangle\rightarrow|3\rangle$, $\Delta_{1}=\omega_{1c}-\omega_{32}$ is the frequency detuning of the coupling laser and the atomic transition $|2\rangle\rightarrow|3\rangle$, $a^{\dag}$($a$) is the creation (annihilation) operator of the cavity photons, $\sigma_{mn}^{j}=|m\rangle\langle n|$ ($m,n=1,2,3$) is the atomic operator, $g=\mu_{13}\sqrt{\omega_{c}/(2\hbar\varepsilon_{0}V)}$ is the cavity-QED coupling coefficient ($\varepsilon_{0}$ is the free space permittivity and $V$ is the cavity mode volume), and $\Omega_1=\mu_{23} E/\hbar$ is the Rabi frequency of the coupling laser ($E$ is the field amplitude and $\mu_{mn}$ is the matrix element of the electric dipole moment).

The dynamics of the system described by Eq. (\ref{equ-Hamiltonian}) can be governed by the quantum Langevin equations as follows \cite{ScullyMOZubairyMS2001,WallsDFandMilburnGJ2007},
\begin{equation}
\begin{split}
	\dot{\rho}=&\frac{1}{i\hbar}[H,\rho]-\frac{1}{2}\{\gamma,\rho\},\\
	\dot a=&[i(\Delta_p-\Delta_{ac})-\frac{(\kappa_{l}+\kappa_{r})}{2} ]a+i g N\rho_{13}+\sqrt{\kappa_{l}/\tau} a_{in,l}\\
	&+\sqrt{\kappa_{r}/\tau} a_{in,r},\label{equ-motion}
\end{split}
\end{equation}
Here $\langle n|\gamma|m\rangle=\gamma_n\delta_{nm}$ defines the relaxation matrix of the atomic energy levels, $\kappa_l$ ($\kappa_r$) is the field decay rate from the left (right) cavity mirror, and $\tau$ is the photon round trip time inside the cavity \cite{Zou2013}.

%\subsection{Analytical solution}
 
To simplify our analysis, we consider a symmetric cavity with $\kappa_l=\kappa_r=\kappa$, and we assume the coupling field is resonant with the atomic transition $|2\rangle\rightarrow|3\rangle$ (i.e. $\Delta_1=0$). In semiclassical approximation, we treat the expectation values of field operators as the corresponding fields \cite{Sawant2016a}, e.g. $\langle{a}\rangle=\alpha$ and $\langle{a^{\dag}}\rangle=\alpha^{*}$. Under the steady-state condition, the time derivatives of the mean values of the system operators vanish, i.e., $\dot{\langle\rho\rangle}=\dot {\langle a\rangle}=0$. Therefore, by solving Eq. (\ref{equ-motion}), the steady-state intracavity field can be attained as follows, 
\begin{widetext}
\begin{equation}
    \alpha=\frac{(\alpha_{in,l}+\alpha_{in,r}) \sqrt{\kappa/\tau}}{\kappa -i (\Delta_p-\Delta_{ac})-\frac{4 i g^2 N \Omega _1^2 \{2 |\alpha|^2 g^2 (2 \gamma _{12} S_{\gamma 1}-i A \Gamma _{32})+ 2 \gamma _{12}^2 \Gamma  S_{\gamma 1}+ \gamma _{12}\Gamma[S_{\gamma 2}-4 i \Omega _1^2 (r-\Gamma _{31})]+\Gamma B X\}}{4 |\alpha|^2 g^2 [\gamma _{12}^2 \Gamma ^2 (\Gamma  \Gamma _{32}+12 \Omega_1^2)+\gamma _{12}(S_{\gamma 3}+2 D \Delta _p^2)+\Gamma (Y+2 C \Omega_1^2)]+\Gamma D[\gamma _{12}^2 (\Gamma ^2+4 \Delta _p^2)+4 \gamma _{12} \Gamma  \Omega _1^2+|X|^2]+o(\alpha)}},\label{equ-intra-field-h}
\end{equation}
\end{widetext}
where $A=\Gamma  r+2 i \Delta _p (\Gamma +r)$, $B=\Gamma _{32} r-2 i \Delta _p (r-\Gamma _{31})$, $C=\Gamma  \Gamma _{32} r+2 \Delta _p^2 (2 \Gamma +r)$, $D=\Gamma  \Gamma _{32} r+4 \Omega _1^2 (\Gamma _{31}+2 r)$, $X=\Gamma  \Delta _p+2 i (\Delta _p^2-\Omega _1^2)$, $Y=\Gamma^2\Gamma _{32} \Delta _p^2+4 \Omega _1^4 (\Gamma +\Gamma _{31}+r)$, $S_{\gamma 1}=(r-\Gamma_{31})(2\Delta_p-i\Gamma)$, $S_{\gamma2}=\Gamma_{32}r(2\Delta_p-i \Gamma)$, $S_{\gamma 3}=\Gamma^2[\Gamma\Gamma_{32}r+2\Omega_1^2(r+2\Gamma)]+2\Gamma(\Gamma_{32}r\Delta_p^2+12\Omega_1^4)$, $o(\alpha)=16 |\alpha|^6  g^6 \Gamma  \Gamma _{32}+16 |\alpha|^4 g^4\gamma _{12} \Gamma   (\Gamma  \Gamma _{32}+6 \Omega _1^2)+4 |\alpha|^4 g^4[\frac{|S_{\gamma 2}|^2}{\Gamma _{32} r}+4 \Gamma  \Omega _1^2 (\Gamma +\Gamma _{32}-r)]$, $\Gamma$ represents the decay rate of the atomic energy level $|3\rangle$ ($\Gamma=\Gamma_{31}+\Gamma_{32}$), $\gamma_{12}$ is the decoherence rate between the ground energy levels $|1\rangle$ and $|2\rangle$, and $r$ is the pumping rate of the incoherent field.

We consider the system is under the collective strong coupling regime with $g\sqrt{N}>(\kappa, \gamma)$ while $g\ll(\kappa, \gamma)$ \cite{RMP87/1379}, as a result, the higher order term of the intracavity field ($o(\alpha)$) can be ignored, and $\alpha$ can be attained by,
\begin{widetext}
\begin{equation}
      \alpha=\frac{(\alpha_{in,l}+\alpha_{in,r}) \sqrt{\kappa/\tau}}{\kappa -i (\Delta_p-\Delta_{ac})-\frac{4 g^2 N \Omega _1^2 (2 A |\alpha|^2 \Gamma_{32} g^2+i \Gamma B X)}{\Gamma  \{4 |\alpha|^2 g^2 [\Gamma^2 \Gamma_{32} \Delta_p^2+2 C \Omega_1^2+4 \Omega_1^4 (\Gamma +\Gamma _{31}+r)]+D|X|^2\}}}.
    \label{equ-intra-field1}
\end{equation}
\end{widetext}
Considering $\gamma_{12}\ll\Gamma$, we have ignored the decoherence rate $\gamma_{12}$ in Eq. (\ref{equ-intra-field-h}) only for attaining simplified analytical solution. While for the numerical results, we treat $\gamma_{12}=0.001\Gamma$. From Eq. (\ref{equ-intra-field1}), it shows that the atomic susceptibility depends on the intracavity field quadratically, which indicates the atomic optical bistability (AOB) of our system \cite{Joshi}. Considering $o(\alpha)$ in Eq. (\ref{equ-intra-field-h}), it will show atomic optical multistability (AOM), in which the atomic susceptibility show higher-order nonlinearility on $\alpha$.

With the input-output relations of a two-sided cavity \cite{WallsDFandMilburnGJ2007,Agarwal2016}, the output field can be attained by,
\begin{equation}
    \alpha_{out,l}=\sqrt{\kappa\tau}\alpha-\alpha_{in,l},\; \alpha_{out,r}=\sqrt{\kappa\tau}\alpha-\alpha_{in,r}.\label{equ-ain-aout}
\end{equation}
For simplicity, we assume $\alpha_{in,r}=\alpha_{in}$ and $\alpha_{in,l}=\alpha_{in}e^{i\varphi}$ where $\varphi$ is the relative phase of two input probe beams. When CPA occurs (i.e., $\varphi=0$ and $\alpha_{out}^l=\alpha_{out}^r=0$), the intensity of the intracavity field $|\alpha_{CPA}|^2$ can be attained as follows,
\begin{widetext}
\begin{equation}
    |\alpha_{CPA}|^2=\frac{4 \Gamma_{31}\Omega _1^2 X_2-r[(\Gamma  \Gamma_{32}+8 \Omega_1^2) (2 g^2 N\Gamma \Delta _p^2-X_2)+8 g^2 N \Omega _1^2 X_3 ]}{4 g^2 \{2 r \Omega _1^2[\Gamma _{32} (\Gamma  \kappa +g^2 N)+2 \kappa  (\Delta _p^2+\Omega _1^2)]+\kappa  X_1\}},\label{equ-thlintra-cpa}
\end{equation}
\end{widetext}
where $X_1=\Gamma _{32} (\Gamma^2 \Delta_p^2-4 \Omega_1^4)+8 \Gamma \Omega _1^2 (\Delta_p^2+\Omega_1^2)$, $X_2=2g^2 N \Gamma \Delta_p^2-\kappa[\Gamma^2 \Delta_p^2+4 (\Delta_p^2-\Omega _1^2)^2]$, $X_3=\Gamma \Delta _p^2+\Gamma_{32}(\Omega_1^2-\Delta_p^2)$. The input intensity for CPA can be derived from $I_{in}=\kappa\tau|\alpha_{CPA}|^2$, and the frequency range ($\Delta_p^T$) of CPA can be attained by solving $|\alpha_{CPA}|^2\geqslant0$. To show clearly the relationship among the intensity, the frequency of the probe field and the pumping rate $r$ (and/or the coupling strength $\Omega_1$) that allows for CPA, we obtain the Taylor series expansion of $|\alpha_{CPA}|^2$ as follows,
\begin{equation}
\begin{split}
    |\alpha_{CPA}|^2_T\approx&\frac{\Gamma_{31} \Omega_1^2 X_2}{g^2 \kappa X_1}\!-\!\frac{r}{4 g^2 \kappa^2 X_1^2}\{8 \Gamma_{31} \Omega_1^4 X_2[2 \kappa  (\Delta_p^2\!+\!\Omega _1^2)\\
    &+\Gamma _{32} (g^2 N+\Gamma  \kappa)]+\kappa  X_1[8 g^2 N \Omega_1^2 X_3\\
    &+(\Gamma  \Gamma _{32}+8 \Omega _1^2)(2 g^2 N \Gamma \Delta_p^2-X_2)]\},\label{equ-thl-Tcpa}
\end{split}
\end{equation}
where the higher-order term of $r$ is negligible in comparison. 

In the following, we verify that our system can degenerate into a two-level system when $\Gamma_{32}=\Omega_1=0$. Although we cannot treat $\Gamma_{32}$ as zero for implementation in a three-level system, we can use an assistant field with a large detuning pumping the population from level $|3\rangle$ into level $|1\rangle$. Therefore, $\Gamma_{32}\approx0$ and $\Gamma_{31}\approx\Gamma$ are attained, and the solution of $\alpha$ is degenerated as follows,
\begin{equation}
    \alpha =\frac{(\alpha_{in,l}+\alpha_{in,r}) \sqrt{\kappa/\tau}}{\kappa -i (\Delta_p-\Delta_{ac})-\frac{2 g^2 N (r-\Gamma)(\Gamma +2 i \Delta_p)}{4 |\alpha|^2 g^2 (2 \Gamma +r)+(\Gamma+2 r)(\Gamma^2+4 \Delta_p^2)}}.\label{equ-intra-field0}
\end{equation}
When the incoherent pump field is absent, the solution is simplified as,
\begin{equation}
    \alpha =\frac{(\alpha_{in,l}+\alpha_{in,r}) \sqrt{\kappa/\tau}}{\kappa -i (\Delta_p-\Delta_{ac})+\frac{2 g^2 N (\Gamma +2 i \Delta_p)}{8 |\alpha|^2 g^2 +\Gamma^2+4 \Delta_p^2}}.\label{equ-tlintra-field}
\end{equation}
It is verified that the solution is consistent with that in reference \cite{Agarwal2016}. However, the distinct advantage of our system is that the manipulated nonlinear dependence of $\alpha$ on the intensity/frequency of the input probe field by $\Omega_1$ or $r$, which can be inferred from Eqs. (\ref{equ-intra-field1}) and (\ref{equ-intra-field0}).

In the same principle, $|\alpha_{CPA}|^2$ of the two-level system can be attained from Eqs. (\ref{equ-ain-aout}) and (\ref{equ-intra-field0}), and can also be attained from Eq. (\ref{equ-thlintra-cpa}) when $\Gamma_{32}=\Omega_1=0$,
\begin{equation}
    |\alpha_{CPA}|^2=\frac{2 g^2 N\Gamma (\Gamma-r)-\kappa (\Gamma +2 r) (\Gamma^2+4 \Delta_p^2)}{4 g^2 \kappa (2 \Gamma +r)}.\label{equ-tlintra-cpa}
\end{equation}
The Taylor series expansion of $|\alpha_{CPA}|^2$ is as follows, 
\begin{widetext}
\begin{equation}
    |\alpha_{CPA}|^2_T=\frac{2 g^2 N \Gamma-\kappa  (\Gamma ^2+4 \Delta _p^2)}{8 g^2 \kappa }-\frac{r [6 g^2 N \Gamma+3 \kappa  (\Gamma ^2+4 \Delta _p^2)]}{16 g^2 \kappa\Gamma},\label{equ-tl-Tcpa}
\end{equation}
\end{widetext}
which can also be attained from the degeneration of Eq. (\ref{equ-thl-Tcpa}).

\section{Numerical discussion}
\begin{figure}[!hbt]
\centering
\includegraphics[width=4cm]{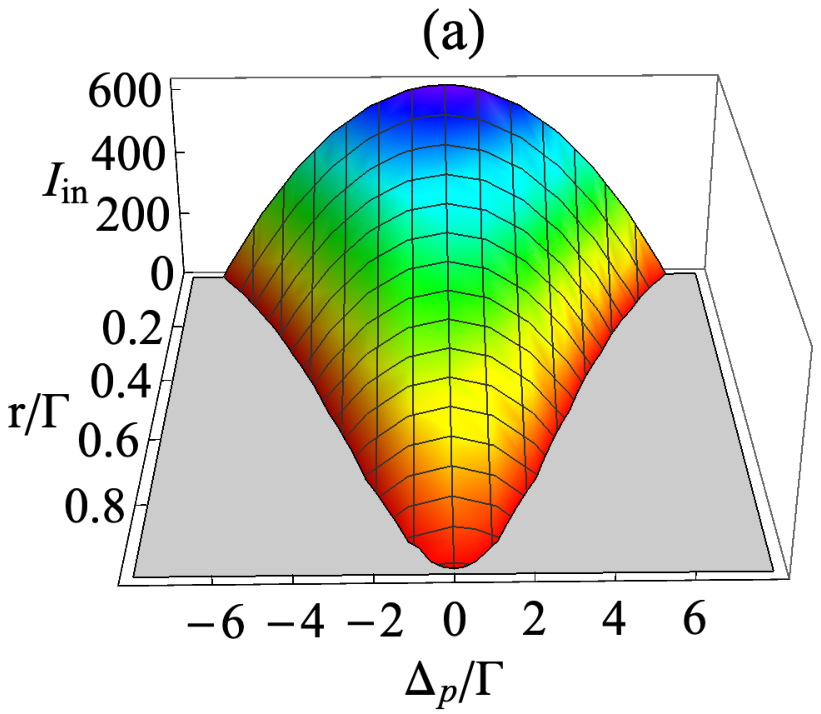}
\includegraphics[width=4cm]{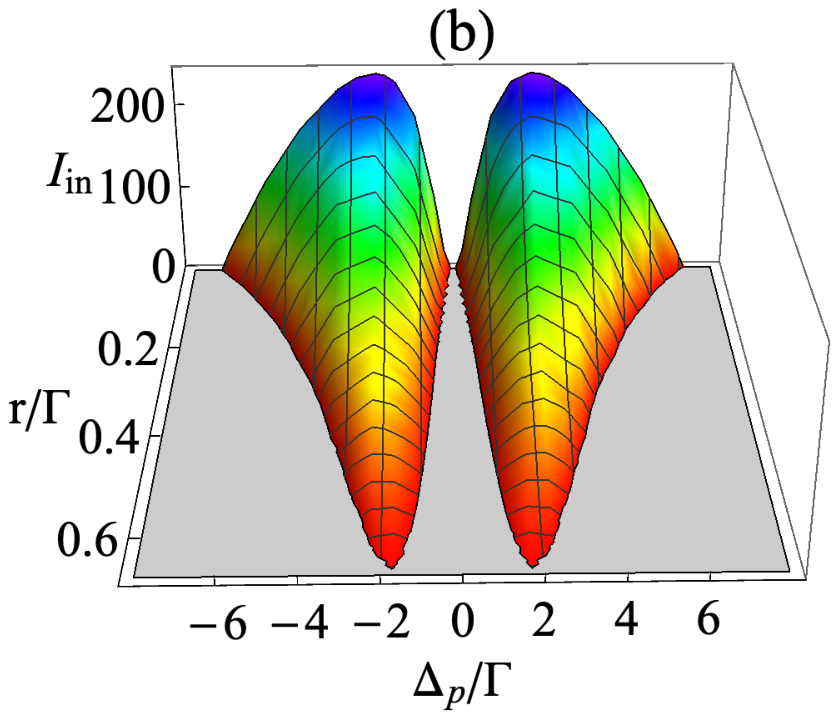}
\includegraphics[width=4cm]{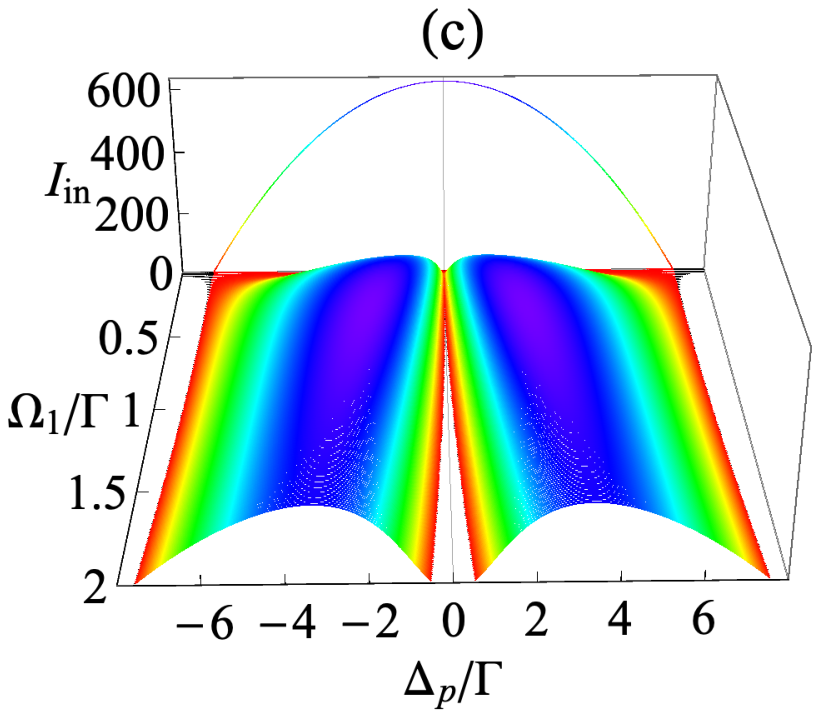}
\includegraphics[width=4cm]{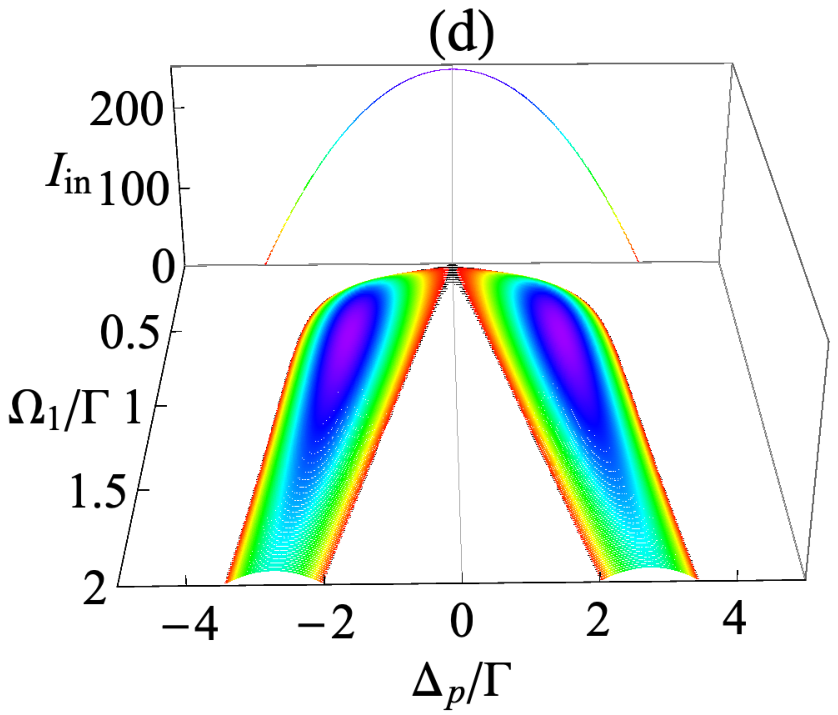}
\caption{The input intensity for CPA ($I_{in}=\kappa\tau|\alpha_{CPA}|^2$) as a function of the frequency of the probe field, the incoherent pumping rate ((a), (b)) and the coupling strength ((c), (d)). (a) $\Omega_1=0$, (b) $\Omega_1=\Gamma$, (c) $r=0$, and (d) $r=0.5\Gamma$.}
\label{fig-cpa-r-o}
\end{figure}

\begin{figure}[!hbt]
\centering
\includegraphics[width=4cm]{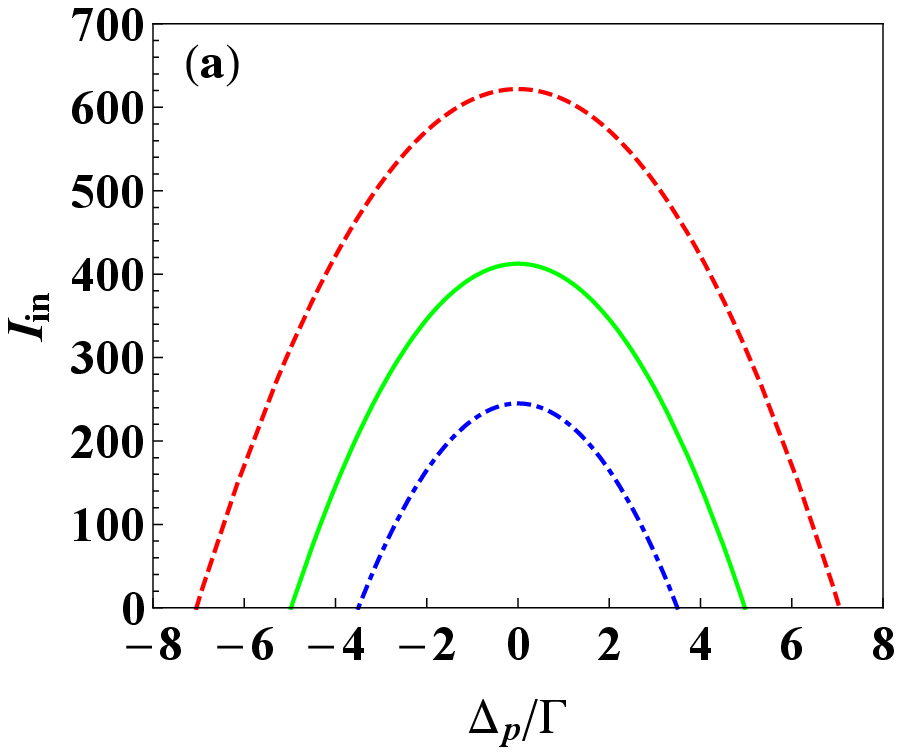}
\includegraphics[width=4cm]{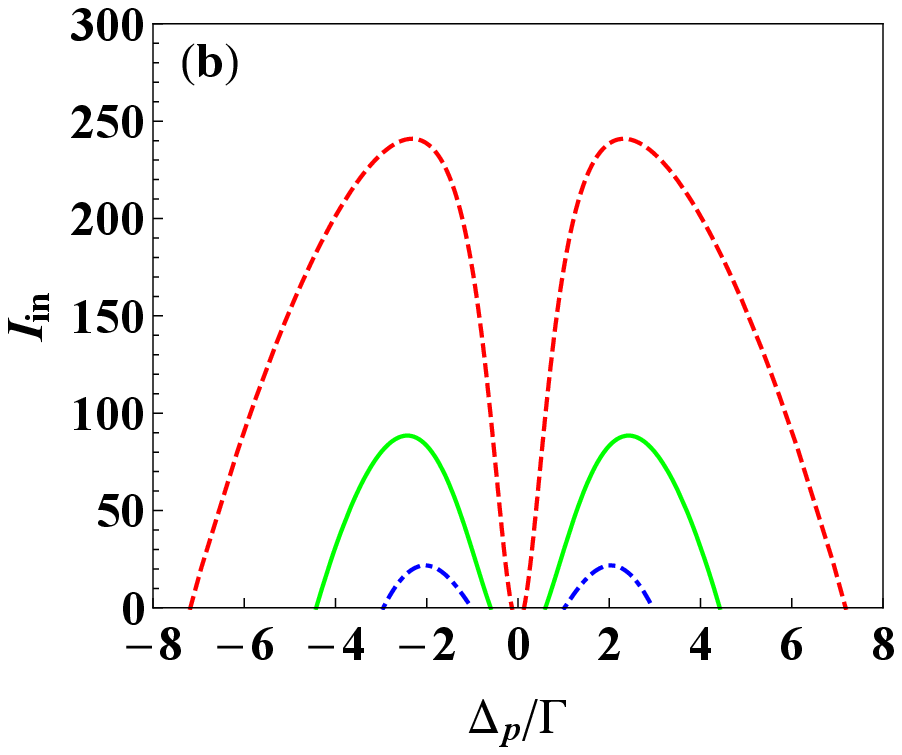}
\caption{The input intensity of CPA versus the frequency detuning of the probe field for (a) $\Omega_1=0$ and (b) $\Omega_1=\Gamma$ with $r=0$ (dashed red line), $r=0.25\Gamma$ (solid green line), and $r=0.5\Gamma$ (dot-dashed blue line).}
\label{fig-Iin-CPA}
\end{figure}

\begin{figure}[!hbt]
    \centering
    \includegraphics[width=4cm]{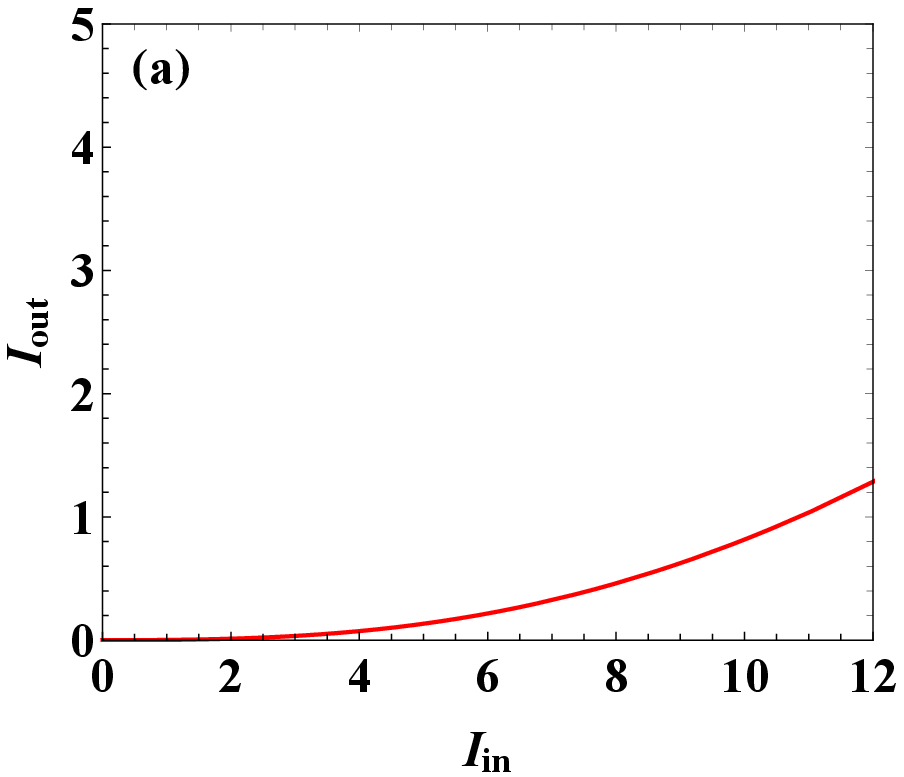}
    \includegraphics[width=4cm]{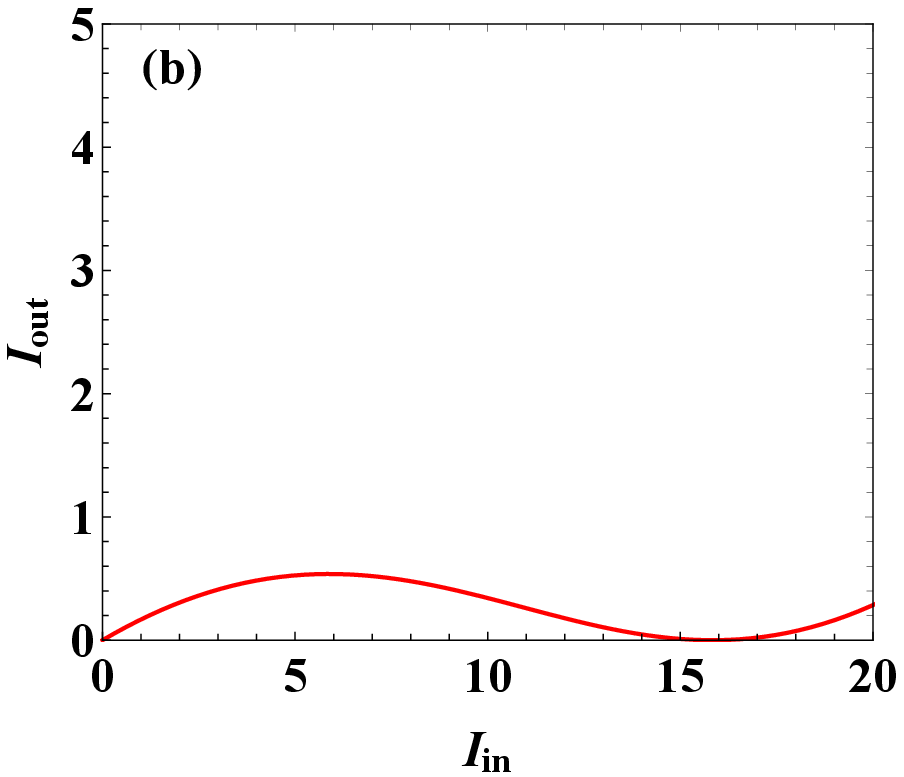}
    \caption{The output intensity versus input intensity for (a) $\Delta_p=7.2\Gamma$ and (b) $\Delta_p=7\Gamma$ with $r=0$ and $\Omega_1=\Gamma$.}
    \label{fig-Iout-Iin}
\end{figure}

\begin{figure*}[!hbt]
\centering
\includegraphics[width=3.5cm]{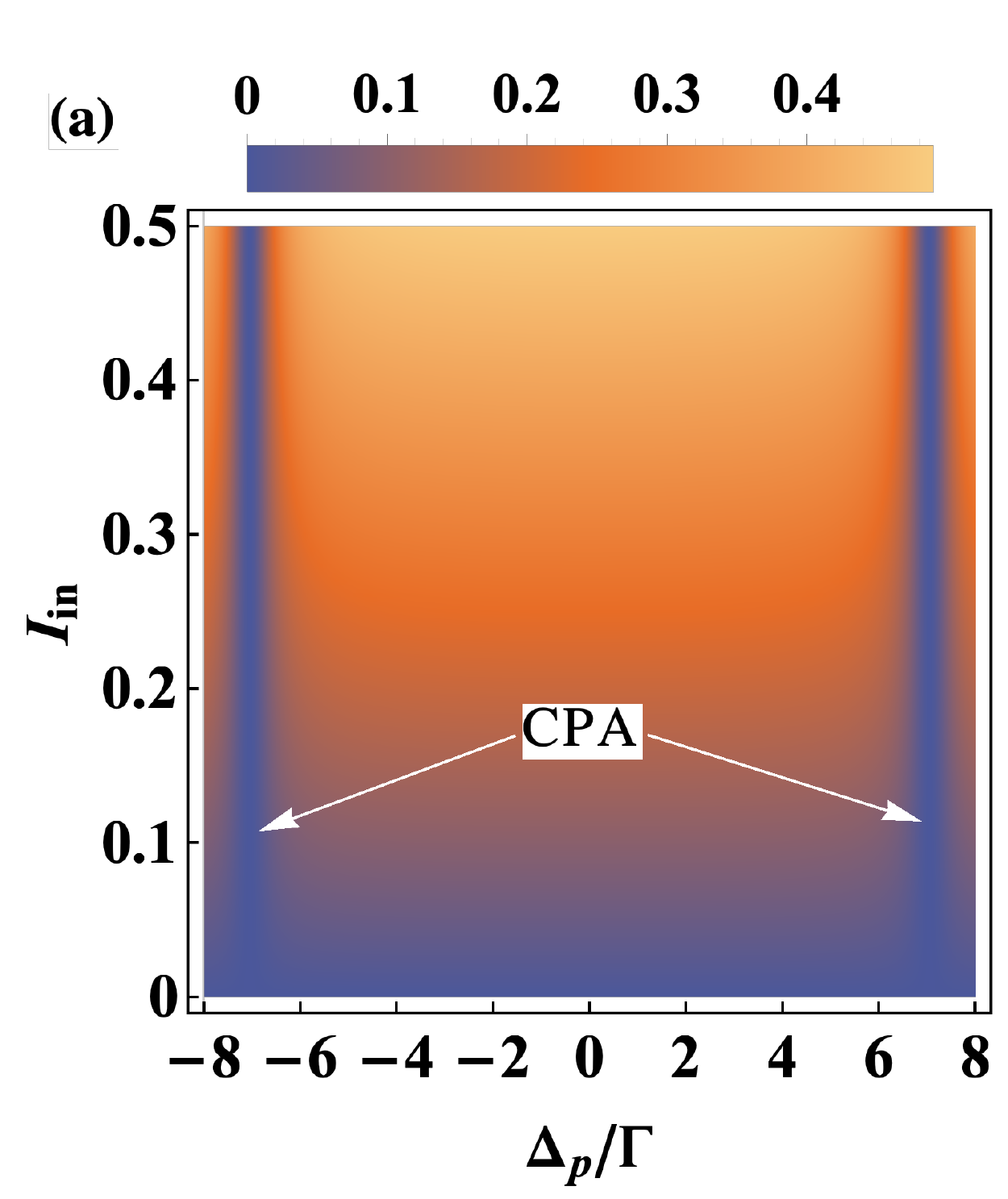}
\includegraphics[width=3.5cm]{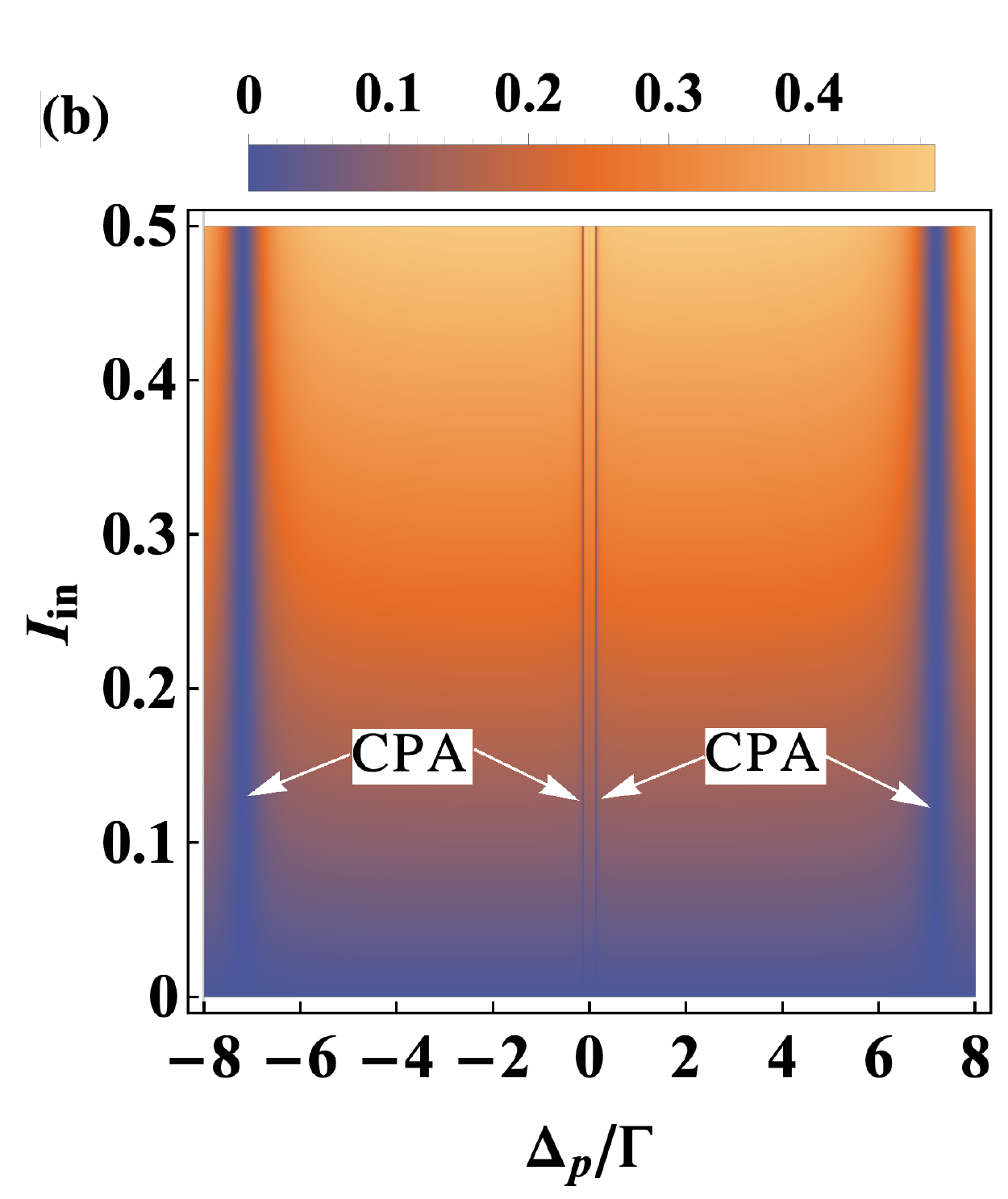}
\includegraphics[width=3.5cm]{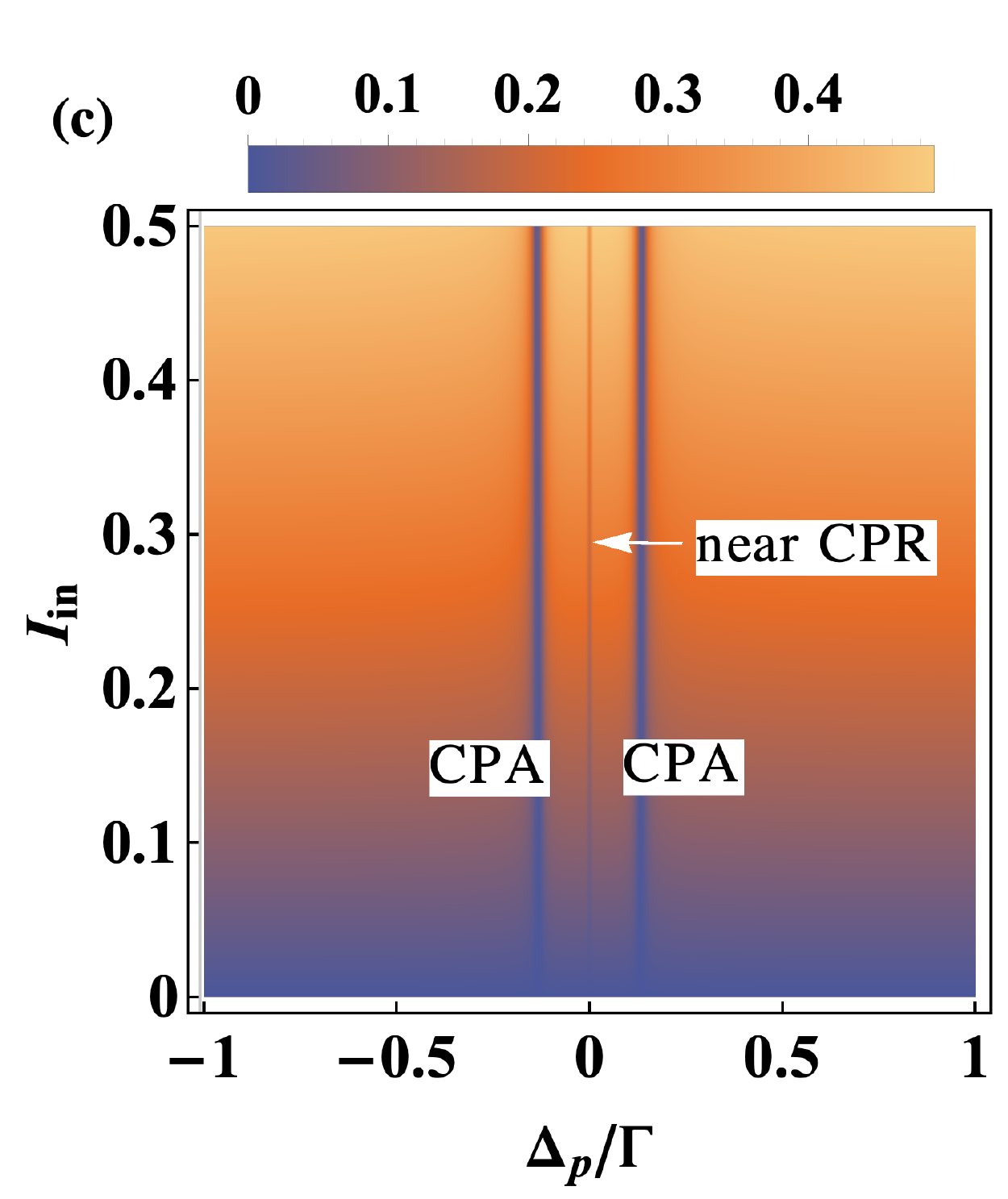}
\caption{The two-dimensional density plot of the output intensity versus the frequency detuning and the input intensity of the probe field in the linear regime with $r=0$. (a) $\Omega_1=0$ and (b), (c) $\Omega_1=\Gamma$ with $r=0$.}\label{fig-cont-CPA}
\end{figure*}

\begin{figure}[!hbt]
	\centering
	\includegraphics[width=4cm]{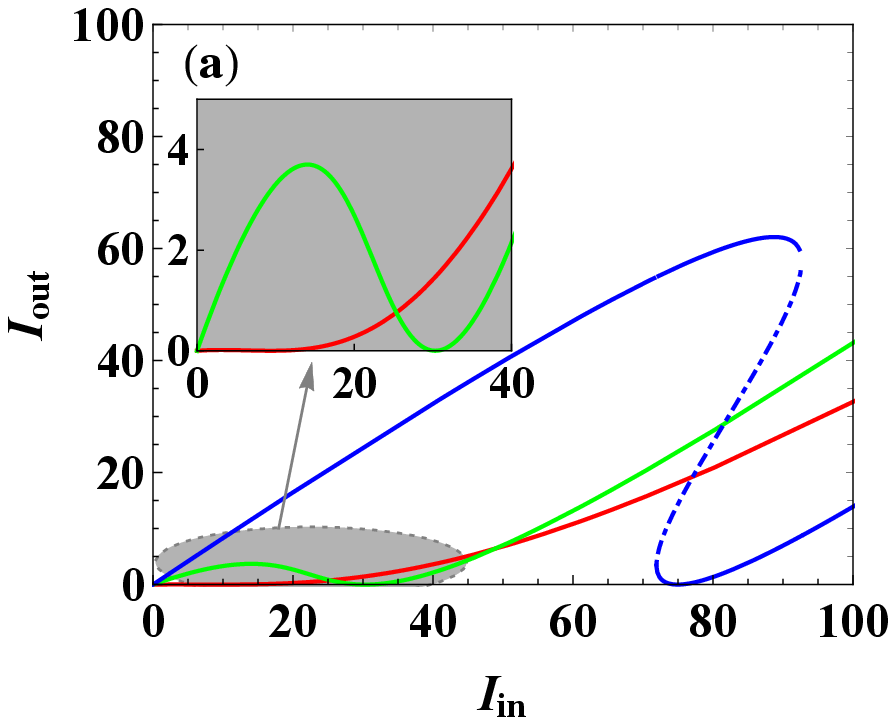}
	\includegraphics[width=4cm]{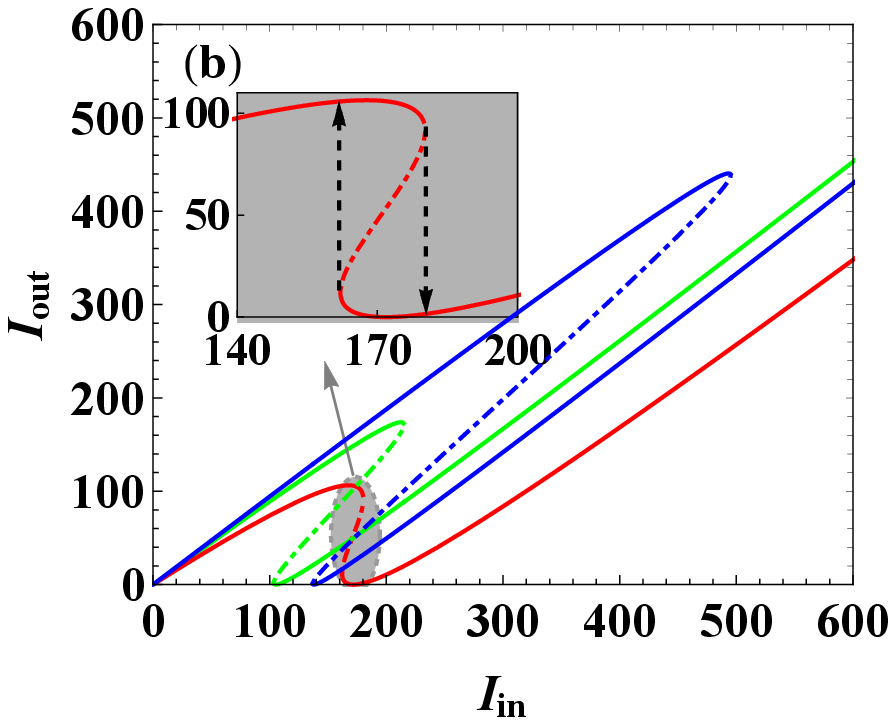}
	\caption{The output intensity versus the input intensity with $r=0$ for (a) $\Delta_p=7\Gamma$ and (b) $\Delta_p=6\Gamma$ with $\Omega_1=0$ (red line), $\Omega_1=1.5\Gamma$ (green line), and $\Omega_1=2.5\Gamma$ (blue line).}\label{fig-Omega-Iout}
\end{figure}

\begin{figure}[!hbt]
	\centering
	\includegraphics[width=4cm]{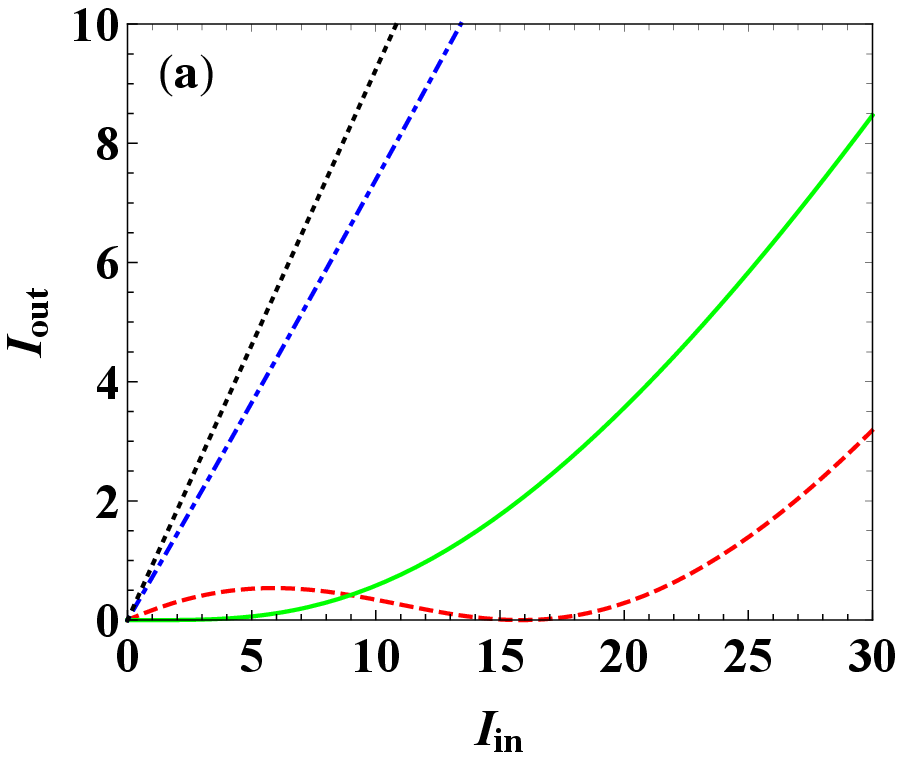}
	\includegraphics[width=4cm]{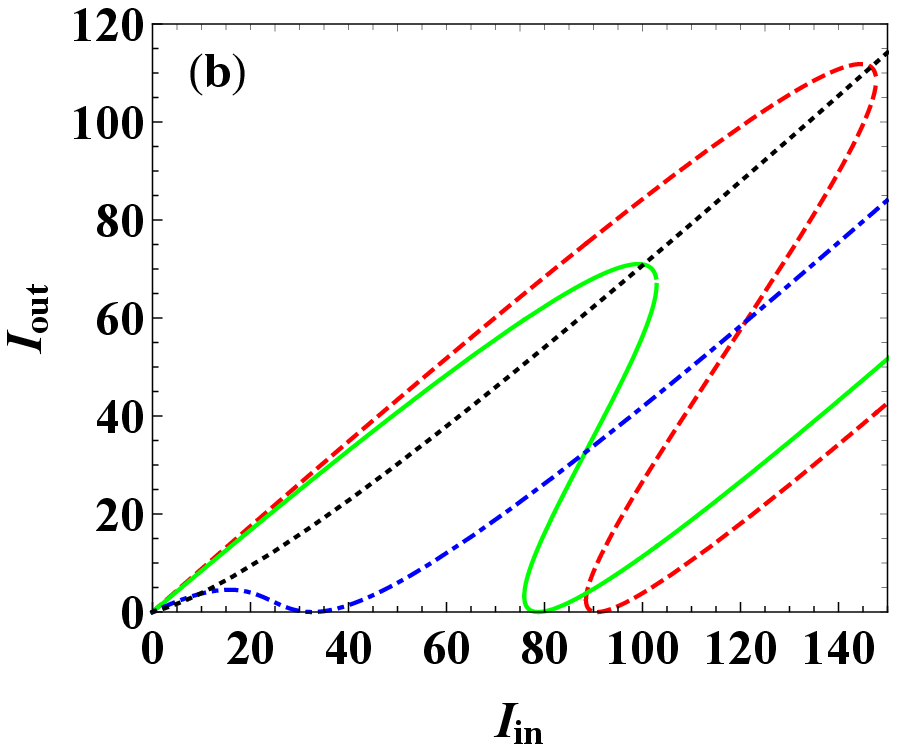}
	\caption{The output intensity versus input intensity for (a) $\Delta_p=7\Gamma$ and (b) $\Delta_p=6\Gamma$ with $r=0$ (dashed red line), $r=0.01\Gamma$ (solid green line), $r=0.05\Gamma$ (dot-dashed blue line), and $r=0.1\Gamma$ (dotted black line).}\label{fig-r-Iout}
\end{figure}

It is inferred from Eqs. (\ref{equ-thl-Tcpa}) and (\ref{equ-tl-Tcpa}) that the intensity of the probe for CPA ($I_{in}(CPA)$) depends on the incoherent pumping rate linearly, and on the coupling strength quartically, which is shown clearly from Fig. \ref{fig-cpa-r-o}. The parameters are $\Delta_1=0$, $g\sqrt{N}=10\Gamma$, $g=0.02\Gamma$, $\kappa\tau=0.01$, $\kappa=\Gamma$, $\gamma_{12}=0.001\Gamma$, and $\Gamma_{31}=\Gamma_{32}=1/2\Gamma$ for Figs. \ref{fig-cpa-r-o}-\ref{fig-r-Iout}.

When $\Omega_1=0$, the system is degenerated to a two-level atom-cavity system, and the input intensity for CPA depends on the frequency ($\Delta_p$) quadratically (see Eq. (\ref{equ-tlintra-cpa}) and Figs. \ref{fig-cpa-r-o}(a), \ref{fig-Iin-CPA}(a)). In addition, the dual-frequency CPA can be attained and tunable by the incoherent pump field. When $r=0$, the analytical and numerical results are consist with that in reference \cite{Agarwal2016} (see Eq. (\ref{equ-tlintra-field})). Increasing the pumping rate, $|\Delta_p|$ is decreasing to maintain $|\alpha_{CPA}|^2\geqslant0$, or lower CPA is attained at a specific $|\Delta_p|$, which can be inferred from Eq. (\ref{equ-tlintra-cpa}). From physical perspective, it can be understood as the decreased absorption of the probe field at the same frequency by the decreased imaginary part of media susceptibility \cite{Xu2012b}. Therefore, the frequency range of CPA is narrowed as shown in Figs. \ref{fig-cpa-r-o}(a) and \ref{fig-Iin-CPA}(a). When the pumping rate is approximately equal to the decay rate of level $|3\rangle$, i.e. $r=0.99\Gamma$, more population will be pumped into level $|3\rangle$, therefore, CPA of the probe laser cannot occur (see Fig. \ref{fig-cpa-r-o}(a)). And it is also calculated that $|\Delta_p^T|=0$ when $r\geqslant0.99\Gamma$ from Eq. (\ref{equ-tlintra-cpa}).

When $\Omega_1\neq0$, the input intensity for CPA depends on the frequency ($\Delta_p$) quartically (see Eq. (\ref{equ-thlintra-cpa})), which induces tunable four-frequency CPA instead of dual-frequency CPA for a certain input intensity within the threshold value of the probe field frequency ($\Delta_p^T$) as shown in Fig. \ref{fig-Iin-CPA}(b) (see also Figs. \ref{fig-cpa-r-o}(b)-\ref{fig-cpa-r-o}(d)). From physical perspective, it can be explained from
the EIT-type interference between two excitation paths $|1\rangle|0\rangle \rightarrow |\Psi_{\pm}\rangle$ which suppresses the polariton excitation and leads to nearly total reflection of the probe fields at $\Delta_p=\Delta_1$. Therefore, CPA cannot be attained at the frequency of two-photon resonance ($\Delta=\Delta_p-\Delta_1=0$). In addition, the intensity for CPA decreases significantly comparing with that in a two-level system, which makes our scheme applicable to weak-field CPA. The above excited polariton states are $\Psi_{\pm}=\frac{1}{\sqrt{2}}[\frac{1}{\sqrt{2}}(\frac{1}{\sqrt{N}}\sum_{j=1}^{N}|1,...3_{j},...1\rangle|0_c\rangle+|1,...1,...1\rangle|1_c\rangle)\pm|2\rangle]$, where $|0_c\rangle$ and $|1_c\rangle$ are zero-photon and one-photon states of the cavity mode.

At the threshold value of $\Delta_p^T$, the cavity-QED system is in the linear excitation regime with low input intensity $I_{in}\leqslant\frac{1}{4}\kappa\tau\Gamma^2/g^2$ \cite{Agarwal2016} (see Fig. \ref{fig-Iout-Iin}(a)), and the linear CPA can be attained as shown in Fig. \ref{fig-cont-CPA}. It is evident that the dual-frequency CPA are attained at $\Delta_p\approx\pm7.0\Gamma$ when $\Omega_1=0$ as shown in Fig. \ref{fig-cont-CPA}(a). When $\Omega_1=\Gamma$, besides the side-band CPA at $\Delta_p\approx\pm7.2\Gamma$, two narrow-band CPA are attained at $\Delta_p\approx\pm0.15\Gamma$ as shown in Fig. \ref{fig-cont-CPA}(b). Figure \ref{fig-cont-CPA}(c) is the partial enlarged drawing of the narrow-band CPA around the resonant frequency of the probe field in Fig. \ref{fig-cont-CPA}(b), in which the near CPR is attained at $\Delta_p=0$ because of the EIT-type destructive interference. Figure \ref{fig-cont-CPA} also shows that the linear CPA, which depends on the frequency of the input probe field, presents the independence on the input intensity.

However, with stronger input probe field or smaller $\Delta_p$, e.g. $I_{in}\geqslant6.25$ or $\Delta_p=7\Gamma$ ($\Delta_p<7\Gamma$ for the two-level system), the cavity-QED system is excited into the nonlinear regime (see Fig. \ref{fig-Iout-Iin}(b)). Different from the linear CPA, the nonlinear CPA depends on both frequency and intensity of the probe field. Although the physical origin of the nonlinear regime is similar for both atom-cavity system, one major advantage of the three-level atom-cavity system is the controllable absorption, dispersion, and nonlinearity of the media by the coupling laser and the pumping field, which provides an approach to exchanging the excitation regime of CPA and realizing tunable nonlinear CPA. In the following, we analyze the dependence of the input-output relation on the coupling strength $\Omega_1$ and the pumping rate $r$, respectively. For comparison, we choose the frequency of the probe field close to and away from the threshold value of $\Delta_p^T$, e.g., $\Delta_p=7\Gamma$ and $\Delta_p=6\Gamma$, respectively.

When $\Delta_p=7\Gamma$, the intracavity field $\alpha$ in Eq. (\ref{equ-intra-field1}) has only one real-value steady-state solution with $\Omega_1=0$, consequently, the output field has single-value solution, and linear CPA is attained as shown by the red curve in Fig. \ref{fig-Omega-Iout}(a). Increasing $\Omega_1$, the strong coupling between levels $|2\rangle$ and $|3\rangle$ enhances the excitation of high-order polaritons which induces the nonlinear regime of the system. However, the multi-value solution of $\alpha$ cannot be attained unless $\Omega_1\geqslant2.23\Gamma$, which can be inferred from Eq. (\ref{equ-intra-field1}). Therefore, the normally nonlinear CPA is attained at $I_{in}\approx30$ with $\Omega_1=1.5\Gamma$ as shown by the green curve in Fig. \ref{fig-Omega-Iout}(a). When $\Omega_1$ is increased as $\Omega_1=2.5\Gamma$, the intracavity field $\alpha$ has three real-value steady-state solutions, and the system is driven into the bistable excitation regime. In one of the bistable branch, CPA occurs as shown by the blue curve in Fig. \ref{fig-Omega-Iout}(a), in which the dot-dashed blue line represents the unstable branch.

When $\Delta_p=6\Gamma$, the intracavity field $\alpha$ has multi-value solution. Therefore, the bistable CPA is attained, although the coupling field is absent, as shown by the red curve in Fig. \ref{fig-Omega-Iout}(b). The black dashed lines with arrow in the inset form the hysteresis cycle of the bistability consisting of two stable branch (solid red line) and an unstable branch (dot-dashed red line). However, when the coupling field is applied, the enhanced Kerr nonlinearity induces wider bistable region as shown by the green and blue curves in Fig. \ref{fig-Omega-Iout}(b). At the same time, increasing the coupling strength decreases the absorption of the probe fields, and even induces the transparency at two-photon resonance \cite{Joshi}. Therefore, it leads to higher input intensity for CPA as shown by the blue curves in Fig. \ref{fig-Omega-Iout}. Nevertheless, a three-level system has distinct advantage of controllable hysteresis cycle of bistable CPA over a two-level system.

In the following discussion, we analyze the manipulation on output field with nonlinear CPA by the incoherent pump field. And the results are presented in Fig. \ref{fig-r-Iout} with $\Omega_1=\Gamma$. It has been inferred that $I_{in}(CPA)$ depends on $r$ negatively linearly (see Eq. (\ref{equ-thl-Tcpa})), therefore, increasing $r$ will decrease $I_{in}(CPA)$ and may change the operation domain of CPA (e.g., from nonlinear CPA to linear CPA as shown by the dashed red and solid green curves in Fig. \ref{fig-r-Iout}(a)). From physical perspective, the controllable nonlinearity by the incoherent pump field can explain the transfer. Increasing the pumping rate may enhance the Kerr nonlinearity \cite{Jafarzadeh2014}, which makes the cavity field easier to reach saturation, therefore, the threshold of OB is decreased (see the dashed red and solid green curves in Fig. \ref{fig-r-Iout}(b)). However, large pumping rate will also destroy the atomic coherence between levels $|1\rangle$ and $|2\rangle$, which reduces the OB region \cite{Gong1997}. With too large pumping rate, more population will be pumped into level $|3\rangle$. However, the population cannot be trapped in level $|3\rangle$, which may lead to the gain of the probe field making the cavity hard to reach the saturation. As a result, the bistable CPA disappears, instead, the normally nonlinear and linear CPA occur as shown by the dot-dashed blue and dotted black curves in Fig. \ref{fig-r-Iout}(b). However, for the frequency near the threshold value of $\Delta_p^T$, e.g. $\Delta_p=7\Gamma$, CPA cannot be attained with too large pumping rate, instead, the near CPR occurs as shown by the dot-dashed blue and dotted black lines in Fig. \ref{fig-r-Iout}(a). This implies a new mechanism of attaining CPR and modifying nonlinear CPA.

\section{Conclusion}
In summary, we have analyzed the manipulation on CPA and CPR with quantum interference and atomic coherence induced by a coherent coupling laser and an incoherent pump field. With the coupling laser, four-frequency CPA instead of dual-frequency CPA is attained in our scheme. And the quartic dependence of $I_{in}(CPA)$ on the coupling strength $\Omega_1$ allows the low-light nonlinear CPA of our scheme, which can also be realized by increasing the pumping rate $r$. Furthermore, the increasing of $r$ reduces the frequency range of CPA and the threshold value of the OB. In comparison, the controllable absorption and nonlinearity in the three-level system provides an approach to exchanging the operation domain between linear CPA and nonlinear CPA without changing the frequency of the probe field. In addition, the CPR or near-CPR is realized in previous studies by changing the relative phase of the input probe fields, however, a new method of attaining near CPR is proposed in our scheme with the controllable linear absorption by $r$. And the results show the possibility of manipulating output state between "0" and "1" with incoherent process, which may have potential applications in quantum information processing.

\begin{acknowledgments}
The author greatly appreciates fruitful discussions with Prof. Jinhui Wu. This work was supported by the Natural Science Foundation of Shaanxi Provincial Department of Education under Grant No. 20JK0682.
\end{acknowledgments}

\end{document}